\pdfoutput=1

\documentclass[12pt,a4paper]{article}

\usepackage{float}
\makeatletter
\let\@tmp\@xfloat     
\usepackage{fixltx2e}
\let\@xfloat\@tmp                    
\makeatother

\usepackage{ifthen} 
\usepackage{longtable} 
\newboolean{pdflatex}
\usepackage{xcolor}
\usepackage{rotating}
\usepackage{multirow}
\setboolean{pdflatex}{true} 

\newboolean{articletitles}
\setboolean{articletitles}{true} 

\newboolean{uprightparticles}
\setboolean{uprightparticles}{false} 

\newboolean{inbibliography}
\setboolean{inbibliography}{false} 

\def\paperauthors{LHCb collaboration} 
\def\paperasciititle{Observation of the B->Jpsi ppbar decays and precision measurements of the Bd/Bs masses}
\def\papertitle{Observation of ${\B^0_{\scaleto{(}{15pt}s\scaleto{)}{15pt}}}\xspace \to J/\psi p \overline{p}$ decays and precision measurements of the ${\B^0_{\scaleto{(}{15pt}s\scaleto{)}{15pt}}}\xspace$ masses}
\def\paperkeywords{{High Energy Physics}, {LHCb}} 
\def\papercopyright{\the\year\ CERN for the benefit of the LHCb collaboration} 
\def\paperlicence{CC-BY-4.0}
\def\paperlicenceurl{https://creativecommons.org/licenses/by/4.0/}

\newboolean{isPRL}
\setboolean{isPRL}{false}

\usepackage{bm}

\usepackage[top=1in, bottom=1.25in, left=1in, right=1in]{geometry}

\usepackage{microtype}
\usepackage{lineno}  
\usepackage{xspace} 
\usepackage{caption} 

\usepackage{float}

\usepackage{graphicx}  
\usepackage{color}
\usepackage{colortbl}
\graphicspath{{./figs/}} 

\usepackage{amsmath} 
\usepackage{amssymb}
\usepackage{amsfonts}
\usepackage{upgreek} 
\usepackage{scalerel}

\newcommand*\patchAmsMathEnvironmentForLineno[1]{%
\expandafter\let\csname old#1\expandafter\endcsname\csname #1\endcsname
\expandafter\let\csname oldend#1\expandafter\endcsname\csname
end#1\endcsname
 \renewenvironment{#1}%
   {\linenomath\csname old#1\endcsname}%
   {\csname oldend#1\endcsname\endlinenomath}%
}
\newcommand*\patchBothAmsMathEnvironmentsForLineno[1]{%
  \patchAmsMathEnvironmentForLineno{#1}%
  \patchAmsMathEnvironmentForLineno{#1*}%
}
\AtBeginDocument{%
\patchBothAmsMathEnvironmentsForLineno{equation}%
\patchBothAmsMathEnvironmentsForLineno{align}%
\patchBothAmsMathEnvironmentsForLineno{flalign}%
\patchBothAmsMathEnvironmentsForLineno{alignat}%
\patchBothAmsMathEnvironmentsForLineno{gather}%
\patchBothAmsMathEnvironmentsForLineno{multline}%
\patchBothAmsMathEnvironmentsForLineno{eqnarray}%
}

\usepackage{hyperxmp}

\usepackage[pdftex,
            pdfauthor={\paperauthors},
            pdftitle={\paperasciititle},
            pdfkeywords={\paperkeywords},
            pdfcopyright={Copyright (C) \papercopyright},
            pdflicenseurl={\paperlicenceurl}]{hyperref}

\usepackage[all]{hypcap} 

\usepackage{xspace} 
\usepackage{upgreek}

\newcommand{\offsetoverline}[2][0.1em]{\kern #1\overline{\kern -#1 #2}}%

\def\lhcb   {\mbox{LHCb}\xspace}

\def\MagUp {\mbox{\em Mag\kern -0.05em Up}\xspace}

\ifthenelse{\boolean{uprightparticles}}%
{

 \def\Pmu         {\ensuremath{\upmu}\xspace}

 \def\Ppi         {\ensuremath{\uppi}\xspace}

 \def\Ppsi        {\ensuremath{\uppsi}\xspace}

 \def\PDelta      {\ensuremath{\Delta}\xspace}                 
 \def\PXi         {\ensuremath{\Xi}\xspace}                 
 \def\PLambda     {\ensuremath{\Lambda}\xspace}                 
 \def\PSigma      {\ensuremath{\Sigma}\xspace}                 
 \def\POmega      {\ensuremath{\Omega}\xspace}                 
 \def\PUpsilon    {\ensuremath{\Upsilon}\xspace}

 \def\PB      {\ensuremath{\mathrm{B}}\xspace}                 
                  
 \def\PD      {\ensuremath{\mathrm{D}}\xspace}

 \def\PJ      {\ensuremath{\mathrm{J}}\xspace}                 
 \def\PK      {\ensuremath{\mathrm{K}}\xspace}

 \def\Pb      {\ensuremath{\mathrm{b}}\xspace}                 
 \def\Pc      {\ensuremath{\mathrm{c}}\xspace}

 \def\Pi      {\ensuremath{\mathrm{i}}\xspace}

 \def\Pp      {\ensuremath{\mathrm{p}}\xspace}

 \def\Ps      {\ensuremath{\mathrm{s}}\xspace}

}
{

 \def\Pmu         {\ensuremath{\mu}\xspace}

 \def\Ppi         {\ensuremath{\pi}\xspace}

 \def\Ppsi        {\ensuremath{\psi}\xspace}                 
                  
 \mathchardef\PDelta="7101
 \mathchardef\PXi="7104
 \mathchardef\PLambda="7103
 \mathchardef\PSigma="7106
 \mathchardef\POmega="710A
 \mathchardef\PUpsilon="7107
                  
 \def\PB      {\ensuremath{B}\xspace}                 
                  
 \def\PD      {\ensuremath{D}\xspace}

 \def\PJ      {\ensuremath{J}\xspace}                 
 \def\PK      {\ensuremath{K}\xspace}

 \def\Pb      {\ensuremath{b}\xspace}                 
 \def\Pc      {\ensuremath{c}\xspace}

 \def\Pi      {\ensuremath{i}\xspace}

 \def\Pp      {\ensuremath{p}\xspace}

 \def\Ps      {\ensuremath{s}\xspace}

}

\makeatletter
\ifcase \@ptsize \relax
  \newcommand{\miniscule}{\@setfontsize\miniscule{4}{5}}
\or
  \newcommand{\miniscule}{\@setfontsize\miniscule{5}{6}}
\or
  \newcommand{\miniscule}{\@setfontsize\miniscule{5}{6}}
\fi
\makeatother

\DeclareRobustCommand{\optbar}[1]{\shortstack{{\miniscule (\rule[.5ex]{1.25em}{.18mm})}
  \\ [-.7ex] $#1$}}

\def\mup        {{\ensuremath{\Pmu^+}}\xspace}
\def\mun        {{\ensuremath{\Pmu^-}}\xspace} 

\def\squark    {{\ensuremath{\Ps}}\xspace}

\def\cquark    {{\ensuremath{\Pc}}\xspace}

\def\bquark    {{\ensuremath{\Pb}}\xspace}

\def\pion   {{\ensuremath{\Ppi}}\xspace}

\def\pip    {{\ensuremath{\pion^+}}\xspace}
\def\pim    {{\ensuremath{\pion^-}}\xspace}

\def\kaon    {{\ensuremath{\PK}}\xspace}
  \def\Kbar    {{\kern 0.2em\overline{\kern -0.2em \PK}{}}\xspace}

\def\KorKbar {\kern 0.18em\optbar{\kern -0.18em K}{}\xspace}

\def\Kp      {{\ensuremath{\kaon^+}}\xspace}
\def\Km      {{\ensuremath{\kaon^-}}\xspace}

  \def\Dbar    {{\kern 0.2em\overline{\kern -0.2em \PD}{}}\xspace}

\def\DorDbar {\kern 0.18em\optbar{\kern -0.18em D}{}\xspace}

\def\B       {{\ensuremath{\PB}}\xspace}
\def\Bbar    {{\ensuremath{\kern 0.18em\overline{\kern -0.18em \PB}{}}}\xspace}

\def\BorBbar    {\kern 0.18em\optbar{\kern -0.18em B}{}\xspace}

\def\Bd      {{\ensuremath{\B^0}}\xspace}
\def\Bs      {{\ensuremath{\B^0_\squark}}\xspace}

\def\Bds     {{\ensuremath{\B_{(\squark)}^0}}\xspace}
\def\Bdsb    {{\ensuremath{\Bbar{}_{(\squark)}^0}}\xspace}

\def\jpsi     {{\ensuremath{{\PJ\mskip -3mu/\mskip -2mu\Ppsi\mskip 2mu}}}\xspace}

\def\Y#1S{\ensuremath{\PUpsilon{(#1S)}}\xspace}

\def\proton      {{\ensuremath{\Pp}}\xspace}
\def\antiproton  {{\ensuremath{\overline \proton}}\xspace}

\def\Lz          {{\ensuremath{\PLambda}}\xspace}

\def\LorLbar     {\kern 0.18em\optbar{\kern -0.18em \PLambda}{}\xspace}

\def\Lc          {{\ensuremath{\Lz^+_\cquark}}\xspace}

\def\Lb           {{\ensuremath{\Lz^0_\bquark}}\xspace}

\def\to                 {\ensuremath{\rightarrow}\xspace}

\newcommand{\mBd}{{\ensuremath{m_{\Bd}}}\xspace}

\newcommand{\mBs}{{\ensuremath{m_{\Bs}}}\xspace}

\def\AT#1     {\ensuremath{A_{\mathrm{T}}^{#1}}\xspace}           

\def\ctl       {\ensuremath{\cos{\theta_\ell}}\xspace}

\def\C#1      {\ensuremath{\mathcal{C}_{#1}}\xspace}                       
\def\Cp#1     {\ensuremath{\mathcal{C}_{#1}^{'}}\xspace}                    
\def\Ceff#1   {\ensuremath{\mathcal{C}_{#1}^{\mathrm{(eff)}}}\xspace}        
\def\Cpeff#1  {\ensuremath{\mathcal{C}_{#1}^{'\mathrm{(eff)}}}\xspace}       
\def\Ope#1    {\ensuremath{\mathcal{O}_{#1}}\xspace}                       
\def\Opep#1   {\ensuremath{\mathcal{O}_{#1}^{'}}\xspace}                    

\newcommand{\aunit}[1]{\ensuremath{\text{\,#1}}}       

\newcommand{\tev}{\aunit{Te\kern -0.1em V}\xspace}
\newcommand{\gev}{\aunit{Ge\kern -0.1em V}\xspace}
\newcommand{\mev}{\aunit{Me\kern -0.1em V}\xspace}
\newcommand{\kev}{\aunit{ke\kern -0.1em V}\xspace}
\newcommand{\ev}{\aunit{e\kern -0.1em V}\xspace}
\newcommand{\mevc}{\ensuremath{\aunit{Me\kern -0.1em V\!/}c}\xspace}
\newcommand{\gevc}{\ensuremath{\aunit{Ge\kern -0.1em V\!/}c}\xspace}
\newcommand{\mevcc}{\ensuremath{\aunit{Me\kern -0.1em V\!/}c^2}\xspace}
\newcommand{\gevcc}{\ensuremath{\aunit{Ge\kern -0.1em V\!/}c^2}\xspace}
\newcommand{\tm}{\aunit{Tm}\xspace}

\def\fb   {\ensuremath{\aunit{fb}}\xspace}
\def\invfb   {\ensuremath{\fb^{-1}}\xspace}

\def\gsim{{~\raise.15em\hbox{$>$}\kern-.85em
          \lower.35em\hbox{$\sim$}~}\xspace}
\def\lsim{{~\raise.15em\hbox{$<$}\kern-.85em
          \lower.35em\hbox{$\sim$}~}\xspace}

\def\sPlot{\mbox{\em sPlot}\xspace}

\def\pt         {\ensuremath{p_{\mathrm{T}}}\xspace}

\def\ptot       {\ensuremath{p}\xspace}

\def\evtgen     {\mbox{\textsc{EvtGen}}\xspace}

\def\geant      {\mbox{\textsc{Geant4}}\xspace}

\def\photos     {\mbox{\textsc{Photos}}\xspace}

\def\pythia     {\mbox{\textsc{Pythia}}\xspace}

\xspace

\def\tell1  {TELL1\xspace}
\def\ukl1   {UKL1\xspace}

\def\pbar        {\kern 0.1em\overline{\kern -0.1em\Pp}{}\xspace}
\def\ppbar     {\ensuremath{\proton\pbar}\xspace}

\def\Bds       {\ensuremath{\B^0_{\scaleto{(}{6.6pt}s\scaleto{)}{6.6pt}}}\xspace}
\def\BJpp      {\ensuremath{\Bds \to \jpsi \ppbar}\xspace}

\def\BsJpp      {\ensuremath{\Bs \to \jpsi \ppbar}\xspace}
\def\BdJpp      {\ensuremath{\Bd \to \jpsi \ppbar}\xspace}

\def\BsJphi      {\ensuremath{\Bs \to \jpsi \phi}\xspace}
\def\BsJKK      {\ensuremath{\Bs \to \jpsi \Kp\Km}\xspace}

\def\thetal    {\ensuremath{\theta_\ell}}

\def\thetav    {\ensuremath{\theta_h}}

\def\ctv       {\ensuremath{\cos{\theta_h}}\xspace}

\usepackage{cite} 
\usepackage{mciteplus}
\usepackage{comment}

\usepackage{longtable} 

\begin{document}

\renewcommand{\thefootnote}{\fnsymbol{footnote}}
\setcounter{footnote}{1}


\begin{titlepage}
\pagenumbering{roman}

\vspace*{-1.5cm}
\centerline{\large EUROPEAN ORGANIZATION FOR NUCLEAR RESEARCH (CERN)}
\vspace*{1.5cm}
\noindent
\begin{tabular*}{\linewidth}{lc@{\extracolsep{\fill}}r@{\extracolsep{0pt}}}
\ifthenelse{\boolean{pdflatex}}
{\vspace*{-2.7cm}\mbox{\!\!\!\includegraphics[width=.14\textwidth]{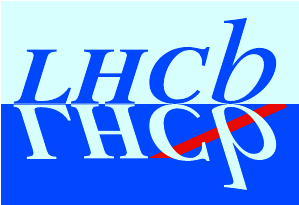}} & &}%
{\vspace*{-1.2cm}\mbox{\!\!\!\includegraphics[width=.12\textwidth]{lhcb-logo.eps}} & &}%
\\
 & & CERN-EP-2019-006 \\  
 & & LHCb-PAPER-2018-046 \\  
 & & \today \\ 
 & & \\
\end{tabular*}

\vspace*{2.0cm}

{\normalfont\bfseries\boldmath\huge
\begin{center}
  \papertitle 
\end{center}
}

\vspace*{2.0cm}

\begin{center}
\paperauthors\footnote{Authors are listed at the end of this paper.}
\end{center}

\vspace{\fill}

\begin{abstract}
\noindent The first observation of the decays $B^0_{(s)} \to J/\psi p \overline{p}$ is reported, using proton-proton collision data corresponding to an integrated luminosity of $5.2{\text{\,fb}}^{-1}$, collected with the \mbox{LHCb}\xspace detector. These decays are suppressed due to limited available phase space, as well as due to Okubo-Zweig-Iizuka or Cabibbo suppression. The measured branching fractions are 
\begin{align*}
\mathcal{B}(B^0 \to J/\psi p \overline{p}) &= (4.51\pm 0.40\; \text{(stat)} \pm 0.44\; \text{(syst)}) \times 10^{-7}, \\
\mathcal{B}(B^0_s \to J/\psi p \overline{p}) &= (3.58\pm 0.19\; \text{(stat)} \pm 0.39\; \text{(syst)}) \times 10^{-6}.
\end{align*}
For the $B^0_s$ meson, the result is much higher than the expected value of $ {\cal O} (10^{-9})$. The small available phase space in these decays also allows for the most precise single measurement of both the $B^0$ mass as ${5279.74 \pm 0.30\; ({\rm stat})\pm 0.10\; ({\rm syst})}\text{\,Me\kern -0.1em V}\xspace$, and the $B^0_s$ mass as ${5366.85 \pm 0.19\; ({\rm stat})\pm 0.13\; ({\rm syst})}\text{\,Me\kern -0.1em V}\xspace$.

\end{abstract}


\begin{center}
  Published in Phys. Rev. Lett. 122, 191804 (2019)
\end{center}

\vspace{\fill}

{\footnotesize 
\centerline{\copyright~\papercopyright, licence \href{\paperlicenceurl}{\paperlicence}.}}
\vspace*{2mm}

\end{titlepage}


\newpage
\setcounter{page}{2}
\mbox{~}
%
%
%
%

\cleardoublepage


\renewcommand{\thefootnote}{\arabic{footnote}}
\setcounter{footnote}{0}



\pagestyle{plain} 
\setcounter{page}{1}
\pagenumbering{arabic}


%

\clearpage

Multiquark hadronic states beyond the well-studied quark-antiquark (meson) and three-quark (baryon) combinations remain elusive even sixty years after their prediction in the quark model~\cite{GellMann:1964nj,Zweig:352337}. Employing an amplitude analysis of $\Lb\to\jpsi\proton\Km$ decays, the LHCb collaboration has found states consistent with $|uudc\overline{c}\rangle$ pentaquarks decaying to $\jpsi p$~\cite{LHCb-PAPER-2015-029,LHCb-PAPER-2016-009} (charge conjugation is implied throughout this Letter). The decays \BJpp are sensitive to pentaquark searches in the $\jpsi p$ and $\jpsi \pbar$ components and to glueball states~\cite{Morningstar:1999rf,Chen:2005mg} in the $p\overline{p}$ system. Baryonic $\Bds$ decays are also interesting to study the dynamics of the final baryon-antibaryon system and its characteristic threshold enhancement, whose underlying origin has still to be completely understood~\cite{Rosner:2003bm}.

\begin{figure}[b]
  \begin{center}
    \includegraphics[width=\linewidth]{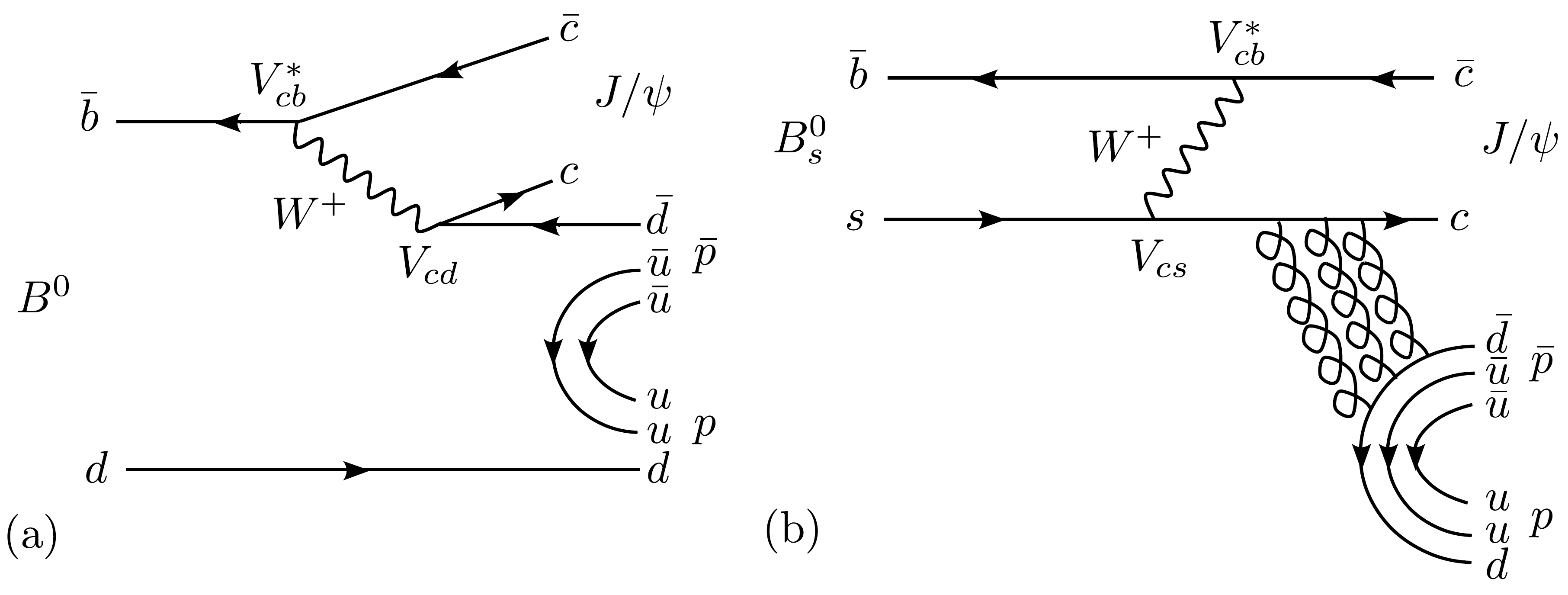}
  \end{center}
  \caption{Leading Feynman diagrams for (a) $\Bd\to \jpsi \ppbar$ and (b) $\Bs\to \jpsi \ppbar$ decays.}
  \label{fig:feyndiag}
\end{figure}

In the leading Feynman diagrams shown in Fig.~\ref{fig:feyndiag}, the $\Bd$ mode is Cabibbo suppressed due to the presence of the Cabibbo-Kobayashi-Maskawa element $V_{cd}$, while the $\Bs$ mode is Okubo-Zweig-Iizuka suppressed~\cite{OKUBO1963165,Zweig:352337,Iizuka:1966}. The na\"ive theoretical expectation for the branching fraction $\mathcal{B}(\BsJpp)$ is at the level of $10^{-9}$~\cite{Hsiao:2014tda}. However, the presence of an intermediate pentaquark or glueball state can enhance the decay rate. The authors of Ref.~\cite{Hsiao:2014tda} pointed out the potential sensitivity of \BsJpp decays to tensor glueball states via a possible resonant contribution of $f_J(2220) \to \proton\antiproton$, which could enhance the \BsJpp decay branching fraction up to order $10^{-6}$. Hints towards such enhancements were noted in a previous LHCb measurement using 1\invfb of $pp$ collision data, where no observation for either mode was made, but a $2.8$ standard deviation excess was seen for the \BsJpp decay~\cite{LHCb-PAPER-2013-029}. 

These decays also allow for high-precision mass measurements. The kinetic energies in the \Bds rest systems of the decay products ($Q$-values) are approximately $306\mev$ for \Bd and $393\mev$ for \Bs decays. The small $Q$-values imply a very small contribution from momentum uncertainties to the \Bds mass measurements.

In this Letter, the first observation of these modes along with their branching fraction and \Bd and \Bs mass measurements are reported employing a data sample corresponding to 5.2\invfb of $pp$ collision data collected by the \lhcb experiment. As a normalization mode, the copious {$\Bs\to \jpsi \phi(\to \Kp\Km)$} sample is used, which is similar in topology to the signal channels.

The \lhcb detector~\cite{Alves:2008zz,LHCb-DP-2014-002} is a single-arm forward
spectrometer covering the \mbox{pseudorapidity} range $2<\eta <5$,
designed for the study of particles containing \bquark or \cquark
quarks. The detector includes a high-precision tracking system
consisting of a silicon-strip vertex detector surrounding the $pp$
interaction region~\cite{LHCb-DP-2014-001}, a large-area silicon-strip detector located
upstream of a dipole magnet with a bending power of about 4\tm, and three stations of silicon-strip detectors and straw
drift tubes~\cite{LHCb-DP-2013-003} placed downstream of the magnet.
The tracking system provides a measurement of the momentum, \ptot, of charged particles with
a relative uncertainty that varies from 0.5\% at low momentum to 1.0\% at 200\gev.\footnote{Natural units with $\hbar = c = 1$ are used throughout.}
Different types of charged hadrons are distinguished using information
from two ring-imaging Cherenkov detectors~\cite{LHCb-DP-2012-003}. 
Muons are identified by a system composed of alternating layers of iron and multiwire proportional chambers~\cite{LHCb-DP-2012-002}.
The online event selection is performed by a trigger~\cite{LHCb-DP-2012-004}, comprising a hardware stage based on information from the muon system, followed by a software stage that applies a full event reconstruction. The software trigger is a combination of event categories mostly relying on identifying \jpsi decays consistent with a \B-meson decay topology with two muon tracks originating from a secondary decay vertex detached from the primary $pp$ collision point.

The $pp$ collision data used in this analysis were collected at center-of-mass energies of 7 and 8\tev ($3$\invfb) and 13\tev ($2.2$\invfb), during the Run~1 (2011 and 2012) and Run~2 (2015 and 2016) run periods, respectively. The data taking conditions differ enough between the two run periods, such that they are analyzed separately and the results combined at the end.

Samples of simulated events are used to study the properties of the signal and control channels.  
The $pp$ collisions are generated using
\pythia~\cite{Sjostrand:2006za,*Sjostrand:2007gs} with a specific \lhcb
configuration~\cite{LHCb-PROC-2010-056}. Decays of hadronic particles are described by \evtgen~\cite{Lange:2001uf}, in which final-state radiation is generated using \photos~\cite{Golonka:2005pn}. For the \BsJphi mode,
simulation samples are generated according to a decay model based on results reported in Ref.~\cite{LHCb-PAPER-2014-059}, while the \BJpp signal modes are generated uniformly in phase space. The interactions of the generated particles with the detector and its response are implemented using the \geant toolkit~\cite{Allison:2006ve, *Agostinelli:2002hh} as described in Ref.~\cite{LHCb-PROC-2011-006}.

The event selection relies on the excellent vertexing and charged particle identification (PID) capabilities of the LHCb detector. For a given particle, the associated primary vertex (PV) corresponds to that with the smallest $\chi^2_{\rm IP}$, defined as the difference in $\chi^2$ between the PV fit including and excluding the particle. Signal candidates are formed starting with a pair of charged tracks, consistent with muons originating from a common vertex significantly displaced from its associated PV and with an invariant mass consistent with the \jpsi meson. Another pair of oppositely charged tracks, identified as protons and originating from a common vertex, is combined with the \jpsi candidate to form a $\Bds$ candidate. The entire decay topology is submitted to a kinematic fit where the dimuon invariant mass is constrained to the known \jpsi mass~\cite{PDG2018}. The \BsJphi control mode candidates are reconstructed in a similar fashion, replacing the $\ppbar$ combination with a pair of charged tracks identified as $\Kp\Km$ candidates, required to have an invariant mass within $\pm 5\mev$ of the known $\phi$-meson mass~\cite{PDG2018}. All charged tracks are required to be of good quality and have \mbox{\pt $>$ 300}~\mev (\mbox{\pt $>$ 550}~\mev) for $p$ or $K$ ($\mu$). For the \BsJphi mode, the contamination from $B^0\to \jpsi K^+\pi^-$ decays with a pion misidentified as a kaon is rejected by imposing a \Bd mass veto and using PID information. At this stage, the combinatorial background dominates, comprising a correctly reconstructed \jpsi meson candidate combined with two unrelated charged tracks.

At this stage, a multidimensional gradient-boosting algorithm~\cite{Rogozhnikov} is used to weight the simulated \BsJphi events to match background-subtracted data distributions in all the training variables. These weights are denoted as GB-weights. The background-subtracted data distributions are obtained using the \sPlot technique~\cite{Pivk:2004ty}. Under the assumption that the relative corrections between data and simulation are similar among different $B^0_{(s)}\to\jpsi h^+ h'^-$ decay topologies, $h^+$ and $h'^-$ being charged hadrons, the GB-weights obtained from the control mode are applied to the signal mode. To validate this assumption, similar GB-weights are derived using another control mode, $B^0\to \jpsi \Kp\pim$, yielding similar results.  

For further background suppression, two multivariate classifiers are applied, each employing a gradient-boosted decision tree (BDT)~\cite{Breiman}. In the first stage, the  BDT$_{\rm kin}$ classifier, based on kinematical and topological variables  of the \Bs candidate, is trained using the \BsJphi decays from simulation as signal proxy, and selected $\jpsi\Kp\Km$ candidates in the mass window $[5450,5700]$\mev as background. For BDT$_{\rm kin}$, only kinematic variables whose distributions are similar between the signal and the control mode are employed. These include the $\ptot$, $\pt$, and $\chi^2_{\rm IP}$ values of the $\Bs$ meson, the $\chi^2$ probability from a kinematic fit~\cite{Hulsbergen:2005pu} to the decay topology, and the impact parameter (IP) of the muons with respect to the associated PV. 

To choose the BDT$_{\rm kin}$ selection cut, the $\BsJpp$ signal figure of merit, $S/\sqrt{S+B}$, is required to exceed five in a 2$\sigma$ window around the \Bs mass peak. The background yield, $B$, is estimated from a fit to the $\jpsi \ppbar$ invariant mass distribution. To estimate the expected signal yield, $S$, the central value of the $\BsJpp$ branching fraction quoted in Ref.~\cite{LHCb-PAPER-2013-029} is used, along with the signal efficiency obtained from simulation.

In the final selection stage, a second classifier, BDT$_{\rm PID}$, uses the hadron PID information from the Cherenkov detector system to distinguish between pions, kaons and protons. Aside from PID, the BDT$_{\rm PID}$ training variables also include the $\ptot$, $\pt$ and $\chi^2_{\rm IP}$ values of the protons. The signal sample is taken as the $\BsJpp$ simulation incorporating the GB-weights for the kinematic variables, while the background sample is taken from events in data with $m(\jpsi p \bar{p})\in [5450,5500]$\mev. The hadron PID variables in the simulation require further corrections to be representative of data. The PID variables are obtained from high-yield calibration samples of $\Lc\to\proton\Km\pip$ and $D^{\ast +}\to D^0(\to K^-\pip)\pi^+$ decays, which can be selected as a function of the $\ptot$, $\pt$ and the number of tracks in the event using only kinematic information~\cite{Aaij:2018vrk}. 
 The optimal BDT$_{\rm PID}$ selection criterion is chosen by maximizing the figure of merit $S/\sqrt{S+B}$, with the initial signal and background yields obtained from a fit to the $m(\jpsi \ppbar)$ distribution after the BDT$_{\rm kin}$ selection.  

For the \BsJphi control mode, the selection is performed using a dedicated classifier, BDT$_{\rm CS}$, which includes the kinematic variables considered in BDT$_{\rm kin}$ with the addition of the PID information. 

After application of all selection requirements, the background is predominantly combinatorial. Approximately 1\% of the selected events contain more than one candidate at this stage; a single candidate is selected randomly. The efficiency of the trigger, detector acceptance, reconstruction and selection procedure is approximately 1\%, as estimated from simulation. 

\begin{figure}
  \begin{center}
    \includegraphics[width=0.7\linewidth]{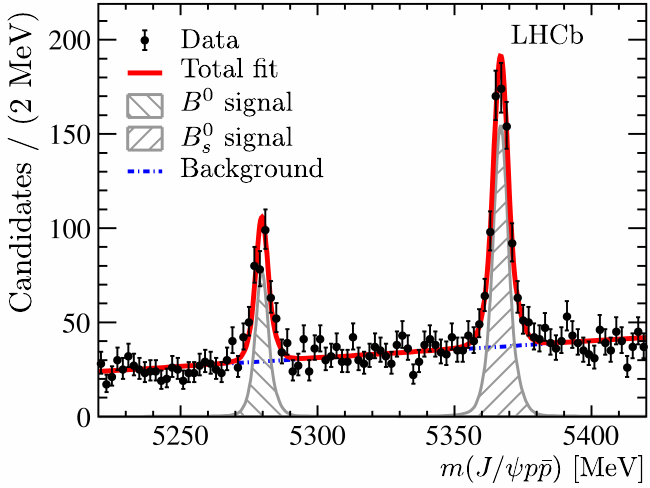}
  \end{center}
  \caption{Fit to the $\jpsi \ppbar$ invariant-mass distribution of the \Bds signal modes.} 
  \label{fig:massFit_signal}
\end{figure}

The \Bd and \Bs signal and background yields are determined via an extended maximum likelihood fit to the $\jpsi\ppbar$ invariant mass distribution in the range $[5220,\;5420]$\mev. Each signal shape is modeled as the sum of two Crystal Ball~\cite{Skwarnicki:1986xj} functions sharing a common peak position, with tails on either sides of the peak to describe the radiative and misreconstruction effects. The background shape is modeled by a first-order polynomial with parameters determined from the fit to data. The signal-model parameters are determined from simulation and only the \Bd and \Bs central mass values are left as free parameters in the fit to data. The detector invariant-mass resolution is in agreement with simulations within a factor of $1.007 \pm 0.004$ as determined with the control mode. Residual discrepancies are accounted for in the systematic uncertainties. In order to validate the fit model, 1000 mass spectra are generated according to the model and fitted employing an alternative model comprising three Gaussian components for the signal and an exponential function for background. The difference between the input value of the yields and the mean of the fitted yields from the alternative model is assigned as a systematic uncertainty. The mass fit for the control mode uses a similar \Bs signal lineshape, with the background modeled by an exponential function. The result of the fit to the combined Run~1 and Run~2 control mode yields a signal of $136{,}800 \pm 400$. The corresponding fit to the signal-mode candidates is shown in Fig.~\ref{fig:massFit_signal} with the results reported in Table~\ref{tab:yields}, where clear signals of \Bd and \Bs are observed.  

\begin{table}[tb]

\caption{Signal yields and masses for \Bd and \Bs mesons.}
\label{tab:yields}
\begin{center}
\begin{tabular}{ c  c  c  c }
\hline
 Mode & Yield & $\Bds$ mass [MeV]\\ 
\hline
 $\BdJpp$ & $256 \pm 22$    & $5279.74\pm0.30$ \\ \hline
 $\BsJpp$ & $609 \pm 31$    & $5366.85\pm0.19$ \\ \hline
\end{tabular}
\end{center}
\end{table}

The branching fractions measured with respect to the ${\Bs\to \jpsi \phi}$ control mode are
\begin{align}
\frac{\mathcal{B}(\BdJpp)}{\mathcal{B}(\Bs\to \jpsi \phi) \times \mathcal{B}(\phi \to \Kp\Km) \times f_s/f_d}  &= \;\frac{N^{\rm corr}_{\BdJpp}}{N^{\rm corr}_{\BsJKK}}, \nonumber \\
\frac{\mathcal{B}(\BsJpp)}{\mathcal{B}(\Bs\to \jpsi \phi) \times \mathcal{B}(\phi \to \Kp\Km)}  &= \;\frac{N^{\rm corr}_{\BsJpp}}{N^{\rm corr}_{\BsJKK}}\nonumber,
\end{align}
where $f_s/f_d$ is the ratio of the \bquark-quark hadronization probabilities into $\Bs$ and $\Bd$ mesons, and $N^{\rm corr}$ denotes efficiency-corrected signal yields. 
For the signal modes, since the physics model is not known {\em a priori}, an event-by-event efficiency correction is applied to the data. It is derived from simulation as a function of the kinematic variables, which are given in detail in the \ifthenelse{\boolean{isPRL}}{Supplemental material~\cite{suppl}.}{Appendix.}
 
Since the control mode has a topology very similar to that of the signal mode, most of the systematic uncertainties cancel in the branching-fraction ratio measurement. Residual systematic effects of the PID efficiency estimation are due to the correction procedure. An alternative PID correction is considered using proton calibration samples from decays of the long-lived $\Lz$ baryon to a proton and a pion, instead of prompt $\Lc$ decays. The difference between the two methods is assigned as a systematic uncertainty. The degree to which the simulation describes hadronic interactions with the detector material is less accurate for baryons than it is for mesons~\cite{Lange:2001uf}. Following Ref.~\cite{LHCb-PAPER-2016-048}, a systematic uncertainty of 4\% (1.1\%) per proton (kaon) is assigned. Other systematic effects include the choice of the fit model, the weighting procedure, the trigger efficiency, and the presence of events with more than one candidate.  The overall systematic uncertainties on the ratio of branching fractions are $7.2\%$ ($7.2\%$) and $6.5\%$ ($6.6\%$) for \Bs (\Bd) meson in Run~1 and Run~2, respectively, where the relevant contributions, listed in Table~\ref{tab:sys}, are added in quadrature. Since the detector and the analysis methods remain the same between the two run periods, the systematic uncertainties are fully correlated while the statistical uncertainties are uncorrelated. The combination of the measurements is taken as a weighted mean to give the branching fraction ratios
\begin{align}
\frac{\mathcal{B}(\BdJpp)}{\mathcal{B}(\Bs\to \jpsi \phi) \times \mathcal{B}(\phi \to \Kp\Km) \times f_s/f_d}  &= ( 0.329 \pm 0.029\; (\text{stat}) \pm 0.022\; (\text{syst}))\times 10^{-2} 
,\nonumber \\
\frac{\mathcal{B}(\BsJpp)}{\mathcal{B}(\Bs\to \jpsi \phi) \times \mathcal{B}(\phi \to \Kp\Km)}  &= ( 0.706 \pm  0.037\; (\text{stat}) \pm 0.048\; (\text{syst}))\times 10^{-2} 
, \nonumber
\end{align}
where the first uncertainty is statistical and the second is systematic. For the absolute branching-fraction determination, the value ${\mathcal{B}(\BsJphi) \times \mathcal{B}(\phi \to \Kp\Km) \times f_s/f_d= (1.314\pm0.016\pm0.079)\times 10^{-4}}$ is obtained from Ref.~\cite{LHCb-PAPER-2012-040} as the product of the two branching ratios, ${\mathcal{B}(\BsJphi) = (10.50\pm0.13\pm0.64) \times 10^{-4}}$ and $\mathcal{B}(\phi \to \Kp \Km) = 0.489 \pm 0.005$, and the ratio of fragmentation probabilities ${f_s/f_d = 0.256 \pm 0.020}$~\cite{fsfd}.
For the \Bs-meson normalization, the updated ratio ${f_s/f_d = 0.259 \pm 0.015}$~\cite{fsfd} is used in Run~1, while for Run~2 it has been multiplied by an additional scale factor of $1.068\pm0.046$~\cite{LHCb-PAPER-2017-001} to take into account the dependence on the center of mass energy. The small $S$-wave $K^+K^-$ fraction under the $\phi(1020)$ resonance, $F_S= 0.0070\pm0.0005$~\cite{LHCb-PAPER-2012-040}, is accounted for as a correction. The absolute branching fractions are then combined to give 
\begin{align*}
\mathcal{B}(\BdJpp)&= (4.51\pm 0.40\; \text{(stat)} \pm 0.44\; \text{(syst)}) \times 10^{-7},\\
\mathcal{B}(\BsJpp)&= (3.58\pm 0.19\; \text{(stat)} \pm 0.39\; \text{(syst)}) \times 10^{-6},
\end{align*}
where the systematic uncertainty is the sum in quadrature of the overall systematic contribution on the ratio of branching fractions, the normalization mode uncertainty and the $f_s/f_d$ uncertainty for the \Bs signal. Table~\ref{tab:sys} summarizes the systematic uncertainties separately for the run periods. The dominant contributions are the normalization, the PID, and the tracking systematic uncertainties. For the \Bd meson, the external normalization measurement from Run~1, $\mathcal{B}(\Bs\to\jpsi \phi) \times \mathcal{B}(\phi \to \Kp\Km)  \times f_s/f_d$~\cite{LHCb-PAPER-2012-040} is used, 
 while for Run~2 the additional energy-dependent correction on $f_s/f_d$ has an uncertainty of 4.3\%. For the \Bs meson, the measured $\mathcal{B}(\Bs\to\jpsi \phi)  \times \mathcal{B}(\phi \to \Kp\Km) \times f_s/f_d$ is divided by $f_s/f_d$ to obtain the \Bs normalization, $\mathcal{B}(\Bs\to\jpsi \phi)  \times \mathcal{B}(\phi \to \Kp\Km) $, resulting in an uncertainty on $f_s/f_d$ independent of the run condition. 
\begin{table}[tb]
\caption{Systematic uncertainties on the branching fraction measurements for Run~1 and Run~2. The total uncertainties on the branching fraction ratios (${\rm BFR}$) are the sum of the systematic uncertainties, added in quadrature. The total uncertainties on the absolute branching fractions ($\mathcal{B}$) include the normalization and the uncertainties on the ratio $f_s/f_d$ from external measurments as well.  
}
\label{tab:sys}
\begin{center}
\begin{tabular}{ c  c  c}
\hline
	& $\mathcal{B}(\BdJpp)$ &  $\mathcal{B}(\BsJpp)$ \\ \cline{2-3}
 	& Run~1 (Run~2) & Run~1 (Run~2) \\ 
\hline
 Fit model & 1.0 (0.5)\%	&  1.0 (0.9)\%  \\ 
 Detector resolution  & 0.6 (0.5)\%  & 0.4 (0.6)\% \\
 PID efficiency  & 5.0 (4.0)\%  & 5.0 (4.0)\%  \\
 Trigger 	& 1.0 (1.0)\%	& 1.0 (1.0)\% \\
 Tracking	& 5.0  (5.0)\%	& 5.0 (5.0)\%  \\
 Simulation weighting	&  0.4 (0.4)\%	& 0.3 (0.3)\%\\
 Multiple candidates & 0.1 (0.1)\%	& 0.1 (0.1)\%	\\
\hline
 Total on ${\rm BFR}$	& 7.2 (6.5)\%	& 	7.2 (6.6)\% \\

\hline
 Normalization   & 6.1 (6.1)\% & 6.1 (6.1)\%	\\
 $f_s/f_d$ &	$ \; - \;\;(4.3)\%$    & 5.8 (5.8)\%	\\
\hline
Total on $\mathcal{B}$ 	& $\;\;9.4 \;(10.1)$\%	& 	11.1 (10.7)\% \\
\hline
\end{tabular}
\end{center}
\end{table}

\begin{table}[b]
\centering
\caption{\label{sys:tab:summassfit} Systematic uncertainties of \Bd and \Bs mass measurements.}
\begin{tabular}{c c c}
\hline
                &       \Bd  & \Bs \\  \cline{2-3} 
                &	[MeV] & [MeV] \\
\hline
	Momentum scale	        & 0.097 & 0.124 \\
	Mass fit model	        & 0.020 & 0.020 \\
	Energy loss correction	& 0.030	& 0.030\\
\hline
	Total			& 0.103 & 0.129	\\
\hline
\end{tabular}
\end{table}

In addition, the small $Q$-values of the \BJpp decays also allow for precise measurements of the \Bd and \Bs masses, with a resolution of 3.3\mev(3.8\mev) for the \Bd (\Bs) meson. The sources of systematic uncertainties include momentum scaling due to imperfections in the magnetic-field mapping derived using well-known narrow resonances, uncertainties on particle interactions with the detector material, and the choice of the signal model, as reported in Table~\ref{sys:tab:summassfit}. The uncertainty on the proton mass is neglected. The final results are 
\begin{align*}
\mBd &= 5279.74 \pm 0.30\; ({\rm stat})\pm 0.10\; ({\rm syst})\mev,\\
\mBs &= 5366.85 \pm 0.19\; ({\rm stat})\pm 0.13\; ({\rm syst})\mev,
\end{align*}
with a correlation of $4\times 10^{-4}$ in the statistical uncertainty. These represent the most precise single measurements for the \Bd and \Bs masses.

In summary, the first observation of the \BdJpp and \BsJpp decays is reported. The measured branching fraction for the \BdJpp decay is consistent with theoretical expectations~\cite{Hsiao:2014tda} while that for \BsJpp is enhanced by two orders of magnitude with respect to predictions without resonant contributions~\cite{Hsiao:2014tda}. More data are needed for glueball and pentaquark searches through a full Dalitz plot analysis. The world's best single measurements of the \Bd and \Bs masses are also reported.

\section*{Acknowledgements}
%
%
\noindent We express our gratitude to our colleagues in the CERN
accelerator departments for the excellent performance of the LHC. We
thank the technical and administrative staff at the LHCb
institutes.
We acknowledge support from CERN and from the national agencies:
CAPES, CNPq, FAPERJ and FINEP (Brazil); 
MOST and NSFC (China); 
CNRS/IN2P3 (France); 
BMBF, DFG and MPG (Germany); 
INFN (Italy); 
NWO (Netherlands); 
MNiSW and NCN (Poland); 
MEN/IFA (Romania); 
MSHE (Russia); 
MinECo (Spain); 
SNSF and SER (Switzerland); 
NASU (Ukraine); 
STFC (United Kingdom); 
NSF (USA).
We acknowledge the computing resources that are provided by CERN, IN2P3
(France), KIT and DESY (Germany), INFN (Italy), SURF (Netherlands),
PIC (Spain), GridPP (United Kingdom), RRCKI and Yandex
LLC (Russia), CSCS (Switzerland), IFIN-HH (Romania), CBPF (Brazil),
 (France), KIT and DESY (Germany), INFN (Italy), SURF (Netherlands),
 PIC (Spain), GridPP (United Kingdom), RRCKI and Yandex
 LLC (Russia), CSCS (Switzerland), IFIN-HH (Romania), CBPF (Brazil),
 PL-GRID (Poland) and OSC (USA).
We are indebted to the communities behind the multiple open-source
software packages on which we depend.
Individual groups or members have received support from Fondazione Fratelli Confalonieri (Italy), 
AvH Foundation (Germany);
EPLANET, Marie Sk\l{}odowska-Curie Actions and ERC (European Union);
ANR, Labex P2IO and OCEVU, and R\'{e}gion Auvergne-Rh\^{o}ne-Alpes (France);
Key Research Program of Frontier Sciences of CAS, CAS PIFI, and the Thousand Talents Program (China);
RFBR, RSF and Yandex LLC (Russia);
GVA, XuntaGal and GENCAT (Spain);
the Royal Society
and the Leverhulme Trust (United Kingdom);
Laboratory Directed Research and Development program of LANL (USA).

\clearpage

{\noindent\normalfont\bfseries\Large Appendix}
\vspace{1cm}

{\noindent\normalfont\bfseries\large A. Efficiency parameterization for the signal mode}\\

\begin{figure}[htb]
\centering
\includegraphics[width=0.55\linewidth]{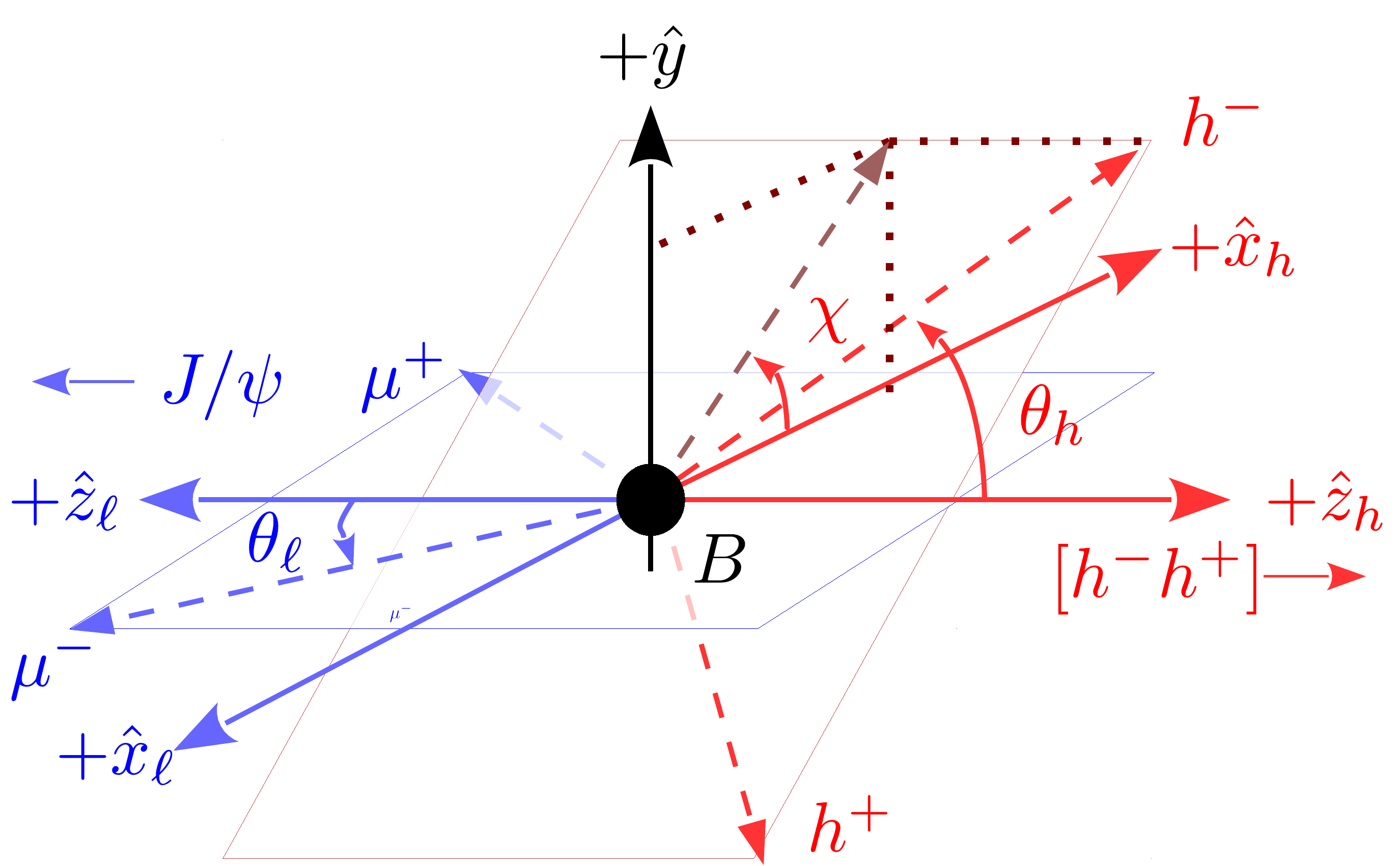}
\caption{The three angular variables $\{\thetal,\thetav,\chi\}$ for the decay $B\to \jpsi (\to \mup\mun) h^+ h^-$, where $h\in \{p,K\}$. The dihadron (in red) and dilepton (in blue) coordinate systems lie back-to-back with a common vertical $\hat{y}$ axis. The angle between the decay planes is $\chi\in (-\pi,\pi]$, while the two helicity angles, $\thetav$ and $\thetal$, are defined in the dihadron and dilepton rest frames, respectively.}
\label{fig:angular_vars}
\end{figure}

\noindent The 4-body phase-space of the decay $B\to \jpsi(\to \mup\mun)h^+h^-$, where $h\in\{p,K\}$, is fully described by four independent kinematic variables. One of them is the dihadron invariant mass $m_{h^+h^-}$. For a given $m_{h^+h^-}$, the topology can be described by three angles, shown in Fig.~\ref{fig:angular_vars}:
\begin{itemize}
\item  $\thetal$ and $\thetav$: the helicity angles defined in the dimuon and dihadron rest frames, respectively;
\item $\chi$: the azimuthal angle between the two decay planes of the dilepton and dihadron systems.
\end{itemize}
Since the final state is self-conjugate, the $h^-$ and the $\mu^-$ particles are chosen to define the angles, for both \Bds and \Bdsb mesons. For the signal mode, the overall efficiency, including trigger, detector acceptance and selection procedure, is obtained from simulation as a function of the four kinematic variables, $\vec{\varphi} \equiv \{m'_{\ppbar}, \cos\theta_{\ell}, \cos\theta_{h}, \chi'\}$. Here, $m'_{\ppbar}$ and $\chi'$ are normalized such that all four variables in $\vec{\varphi}$ lie in the range $(-1,1]$. The efficiency is parameterized as the product of Legendre polynomials
\begin{equation}
\label{eqn:eff_param_def}
\varepsilon(\vec{\varphi}) = \sum_{i,j,k,l} c_{i,j,k,l}~P(\cos\theta_{\ell},i)P(\cos\theta_{h},j)P(\chi',k)P(m'_{\ppbar},l)\nonumber,
\end{equation}
where $P(x,n)$ is a Legendre polynomial of order $n$ in $x\in(-1,1]$. Employing the maximum order of the polynomials as $\{3,7,7,5\}$ for $\{m'_{\ppbar}, \cos\theta_{\ell}, \cos\theta_{h}, \chi'\}$, respectively, was found to give a good parameterization. Simulation samples are employed, where $\BJpp$ events are generated uniformly in phase space. The coefficients, $c_{i,j,k,l}$, are determined from the simulation using a moments technique employing the orthogonality of Legendre polynomials
\begin{align*}
\label{eqn:eff_coeff_mom}
      c_{i,j,k,l}\; =\; & 
                        C \sum_{n=0}^{N_{\rm recon}} 
                        \left(\frac{2i+1}{2}\right)\left(\frac{2j+1}{2}\right)\left(\frac{2k+1}{2}\right)\left(\frac{2l+1}{2}\right) \\\nonumber
 & \hspace{0.3cm}\times P(\ctl,i)P(\ctv,j)P(\chi',k) P(m'_{\ppbar},l).
\end{align*}
The sum is over the number of reconstructed decays, $N_{\rm recon}$, in the simulation sample after all selection criteria. The prefactor $C$ ensures appropriate normalization. For a given data candidate, the corresponding kinematic variables, $\vec{\varphi}$, are reconstructed and the efficiency, $\varepsilon(\vec{\varphi})$, is computed according to the parameterization. The candidate is subsequently assigned a weight, $1/\varepsilon(\vec{\varphi})$, to account for the detector efficiency.\\

{\noindent\normalfont\bfseries\large B. Combining the Run~1 and Run~2 branching fraction results}\\

\noindent The ratio of branching fractions for Run~1 and Run~2 are provided here separately. For Run~1, 
\begin{align*}
\frac{\mathcal{B}(\BdJpp)}{\mathcal{B}(\Bs\to \jpsi \phi)\times \mathcal{B}(\phi \to \Kp\Km)  \times f_s/f_d}  &=  (0.35 \pm 0.05 ({\rm stat})\pm 0.02 ({\rm syst}))\times 10^{-2}, \nonumber \\
\frac{\mathcal{B}(\BsJpp)}{\mathcal{B}(\Bs\to \jpsi \phi)\times \mathcal{B}(\phi \to \Kp\Km) }  &=(0.68 \pm 0.06 ({\rm stat})\pm 0.05 ({\rm syst}))\times 10^{-2}, \nonumber 
\end{align*}
while for Run~2,
\begin{align*}
\frac{\mathcal{B}(\BdJpp)}{\mathcal{B}(\Bs\to \jpsi \phi)\times \mathcal{B}(\phi \to \Kp\Km) \times f_s/f_d}  &=  (0.32 \pm 0.04 ({\rm stat})\pm 0.02 ({\rm syst}))\times 10^{-2}. \nonumber \\
\frac{\mathcal{B}(\BsJpp)}{\mathcal{B}(\Bs\to \jpsi \phi)\times \mathcal{B}(\phi \to \Kp\Km) }  &=(0.72 \pm 0.05 ({\rm stat}) \pm 0.05({\rm syst}))\times 10^{-2}. \nonumber
 \end{align*}
The absolute branching fractions obtained from our present knowledge of ${\mathcal{B}(\Bs\to \jpsi \phi) \times \mathcal{B}( \phi \to \Kp\Km)\times f_s/f_d} =(1.314\pm0.016\pm0.079)\times 10^{-4} $ and ${f_s/f_d =0.259 \pm 0.015}$ are, for Run~1,
\begin{align*}
\mathcal{B}(\BdJpp)&= (4.54 \pm 0.62 ({\rm stat}) \pm 0.33 ({\rm syst}) \pm  0.28 ({\rm norm}))\times 10^{-7}, \\ 
\mathcal{B}(\BsJpp)&= (3.45 \pm  0.31 ({\rm stat}) \pm 0.25 ({\rm syst}) \pm  0.21 ({\rm norm})\pm  0.20 (f_s/f_d)) \times 10^{-6},
\end{align*}
and for Run~2,
\begin{align*}
\mathcal{B}(\BdJpp)&= (4.49 \pm 0.52 ({\rm stat}) \pm 0.29 ({\rm syst})  \pm 0.28 ({\rm norm}) \pm 0.19 (f_s/f_d) ) \times 10^{-7}, \\ 
\mathcal{B}(\BsJpp)&= (3.66 \pm 0.25({\rm stat}) \pm 0.24 ({\rm syst}) \pm  0.22 ({\rm norm})\pm  0.21 (f_s/f_d)) \times 10^{-6}.
\end{align*}
To combine the results from the Run~1 and Run~2 data samples, the systematic uncertainties are taken as fully correlated between the run periods, since the data are collected employing the same detector and analysis techniques. The statistical uncertainties are considered uncorrelated since the datasets are disjoint. The covariance matrices 
are constructed as
\begin{align*}
V =
\begin{pmatrix}
\sigma_{\text{stat, Run~1}}^2 + \sigma_{\text{syst, Run~1}}^2 &  \sigma_{\text{syst, Run~1}} \times \sigma_{\text{syst, Run~2}} \\
\sigma_{\text{syst, Run~1}} \times \sigma_{\text{syst, Run~2}} & \sigma_{\text{stat, Run~2}}^2 + \sigma_{\text{syst, Run~2}}^2
\end{pmatrix},
\end{align*}
with the numerical values 
for the absolute branching fraction combination as
\begin{align*}
V_{\Bd} &=
\begin{pmatrix}
 0.5667  & 0.1916  \\
 0.1916  & 0.4722  \\
\end{pmatrix}\times 10^{-14},\\
V_{\Bs} &=
\begin{pmatrix}
  0.2401   & 0.1502   \\
  0.1502   & 0.2142   \\
\end{pmatrix} \times 10^{-12}.
\end{align*}
The weighted mean value and uncertainty are then calculated as
\begin{align*}
\overline{x} &= \sum_{i\in\{1,2\}} w_i x_i, \\
\sigma_x^2 &= \sum_{i,k\in\{1,2\}} w_i  w_k  V_{ik},
\end{align*}
respectively, where the variable $x$ denotes the branching fraction and the weights are obtained from the inverse of the aforementioned covariance matrix as
\begin{align*}
w_i = \displaystyle \frac{\displaystyle \sum_{k\in\{1,2\}} (V^{-1})_{ik}}{\displaystyle \sum_{j,k\in\{1,2\}} (V^{-1})_{jk}}.
\end{align*}\\


\addcontentsline{toc}{section}{References}
\setboolean{inbibliography}{true}
\bibliographystyle{LHCb}
\bibliography{main,LHCb-PAPER,LHCb-CONF,LHCb-DP,LHCb-TDR,myref,biblio}

\newpage

\newpage
\centerline
{\large\bf LHCb Collaboration}
\begin
{flushleft}
\small
R.~Aaij$^{29}$,
C.~Abell{\'a}n~Beteta$^{46}$,
B.~Adeva$^{43}$,
M.~Adinolfi$^{50}$,
C.A.~Aidala$^{77}$,
Z.~Ajaltouni$^{7}$,
S.~Akar$^{61}$,
P.~Albicocco$^{20}$,
J.~Albrecht$^{12}$,
F.~Alessio$^{44}$
% ...

{\footnotesize \it

% ...
$ ^{7}$Universit{\'e} Clermont Auvergne, CNRS/IN2P3, LPC, Clermont-Ferrand, France\\
% ...
$ ^{12}$Fakult{\"a}t Physik, Technische Universit{\"a}t Dortmund, Dortmund, Germany\\
% ...
$ ^{20}$INFN Laboratori Nazionali di Frascati, Frascati, Italy\\
% ...
$ ^{29}$Nikhef National Institute for Subatomic Physics, Amsterdam, Netherlands\\
% ...
$ ^{43}$Instituto Galego de F{\'\i}sica de Altas Enerx{\'\i}as (IGFAE), Universidade de Santiago de Compostela, Santiago de Compostela, Spain\\
$ ^{44}$European Organization for Nuclear Research (CERN), Geneva, Switzerland\\
% ...
$ ^{46}$Physik-Institut, Universit{\"a}t Z{\"u}rich, Z{\"u}rich, Switzerland\\
% ...
$ ^{50}$H.H. Wills Physics Laboratory, University of Bristol, Bristol, United Kingdom\\
% ...
$ ^{61}$University of Cincinnati, Cincinnati, OH, United States\\
% ...
$ ^{63}$Syracuse University, Syracuse, NY, United States\\
% ...
$ ^{77}$University of Michigan, Ann Arbor, United States, associated to $^{63}$\\
% ...
}
\end{flushleft}

\clearpage

\end{document}